\documentclass{webofc}
\usepackage[varg]{txfonts}   
\usepackage{calrsfs}
\begin{document}
\title{Galactic star formation with NIKA2 (GASTON):\\
Filament convergence and its link to star formation}
%
%

\author{\firstname{N.} \lastname{Peretto}\inst{\ref{Cardiff}}\fnsep\thanks{\email{Nicolas.Peretto@astro.cf.ac.uk}} 
	 \and \firstname{R.}~\lastname{Adam} \inst{\ref{LLR}}
  \and  \firstname{P.}~\lastname{Ade} \inst{\ref{Cardiff}}
  \and  \firstname{H.}~\lastname{Ajeddig} \inst{\ref{CEA}}
  \and  \firstname{P.}~\lastname{Andr\'e} \inst{\ref{CEA}}
  \and \firstname{E.}~\lastname{Artis} \inst{\ref{LPSC}}
  \and  \firstname{H.}~\lastname{Aussel} \inst{\ref{CEA}}
  \and \firstname{A.}~\lastname{Bacmann}\inst{\ref{IPAG}}
  \and  \firstname{A.}~\lastname{Beelen} \inst{\ref{IAS}}
  \and  \firstname{A.}~\lastname{Beno\^it} \inst{\ref{Neel}}
  \and  \firstname{S.}~\lastname{Berta} \inst{\ref{IRAMF}}
  \and  \firstname{L.}~\lastname{Bing} \inst{\ref{LAM}}
  \and  \firstname{O.}~\lastname{Bourrion} \inst{\ref{LPSC}}
  \and  \firstname{M.}~\lastname{Calvo} \inst{\ref{Neel}}
  \and  \firstname{A.}~\lastname{Catalano} \inst{\ref{LPSC}}
  \and  \firstname{M.}~\lastname{De~Petris} \inst{\ref{Roma}}
  \and  \firstname{F.-X.}~\lastname{D\'esert} \inst{\ref{IPAG}}
  \and  \firstname{S.}~\lastname{Doyle} \inst{\ref{Cardiff}}
  \and  \firstname{E.~F.~C.}~\lastname{Driessen} \inst{\ref{IRAMF}}
  \and  \firstname{A.}~\lastname{Gomez} \inst{\ref{CAB}}
  \and  \firstname{J.}~\lastname{Goupy} \inst{\ref{Neel}}
  \and  \firstname{F.}~\lastname{K\'eruzor\'e} \inst{\ref{LPSC}}
  \and  \firstname{C.}~\lastname{Kramer} \inst{\ref{IRAME}}
  \and  \firstname{B.}~\lastname{Ladjelate} \inst{\ref{IRAME}}
  \and  \firstname{G.}~\lastname{Lagache} \inst{\ref{LAM}}
  \and  \firstname{S.}~\lastname{Leclercq} \inst{\ref{IRAMF}}
  \and  \firstname{J.-F.}~\lastname{Lestrade} \inst{\ref{LERMA}}
  \and  \firstname{J.-F.}~\lastname{Mac\'ias-P\'erez} \inst{\ref{LPSC}}
  \and  \firstname{A.}~\lastname{Maury} \inst{\ref{CEA}}
  \and  \firstname{P.}~\lastname{Mauskopf} \inst{\ref{Cardiff},\ref{Arizona}}
  \and \firstname{F.}~\lastname{Mayet} \inst{\ref{LPSC}}
  \and  \firstname{A.}~\lastname{Monfardini} \inst{\ref{Neel}}
  \and  \firstname{M.}~\lastname{Mu\~noz-Echeverr\'ia} \inst{\ref{LPSC}}
  \and  \firstname{L.}~\lastname{Perotto} \inst{\ref{LPSC}}
  \and  \firstname{G.}~\lastname{Pisano} \inst{\ref{Cardiff}}
  \and  \firstname{N.}~\lastname{Ponthieu} \inst{\ref{IPAG}}
  \and  \firstname{V.}~\lastname{Rev\'eret} \inst{\ref{CEA}}
 \and 	\firstname{A.} \lastname{Rigby}\inst{\ref{Cardiff}} 
 \and \firstname{I.} \lastname{Ristorcelli}\inst{\ref{IRAP}}
  \and  \firstname{A.}~\lastname{Ritacco} \inst{\ref{IAS}, \ref{ENS}}
  \and  \firstname{C.}~\lastname{Romero} \inst{\ref{Pennsylvanie}}
  \and  \firstname{H.}~\lastname{Roussel} \inst{\ref{IAP}}
  \and  \firstname{F.}~\lastname{Ruppin} \inst{\ref{MIT}}
  \and  \firstname{K.}~\lastname{Schuster} \inst{\ref{IRAMF}}
  \and  \firstname{S.}~\lastname{Shu} \inst{\ref{Caltech}}
  \and  \firstname{A.}~\lastname{Sievers} \inst{\ref{IRAME}}
  \and  \firstname{C.}~\lastname{Tucker} \inst{\ref{Cardiff}}
  \and  \firstname{R.}~\lastname{Zylka} \inst{\ref{IRAMF}}
}

\institute{
  School of Physics and Astronomy, Cardiff University, Queen’s Buildings, The Parade, Cardiff, CF24 3AA, UK 
  \label{Cardiff}
  \and
  LLR (Laboratoire Leprince-Ringuet), CNRS, École Polytechnique, Institut Polytechnique de Paris, Palaiseau, France
  \label{LLR}
  \and
  AIM, CEA, CNRS, Universit\'e Paris-Saclay, Universit\'e Paris Diderot, Sorbonne Paris Cit\'e, 91191 Gif-sur-Yvette, France
  \label{CEA}
  \and
  Univ. Grenoble Alpes, CNRS, Grenoble INP, LPSC-IN2P3, 53, avenue des Martyrs, 38000 Grenoble, France
  \label{LPSC}
  \and
  Institut d'Astrophysique Spatiale (IAS), CNRS, Universit\'e Paris Sud, Orsay, France
  \label{IAS}
  \and
  Institut N\'eel, CNRS, Universit\'e Grenoble Alpes, France
  \label{Neel}
  \and
  Institut de RadioAstronomie Millim\'etrique (IRAM), Grenoble, France
  \label{IRAMF}
  \and
  Aix Marseille Univ, CNRS, CNES, LAM (Laboratoire d'Astrophysique de Marseille), Marseille, France
  \label{LAM}
  \and 
  Dipartimento di Fisica, Sapienza Universit\`a di Roma, Piazzale Aldo Moro 5, I-00185 Roma, Italy
  \label{Roma}
  \and
  Univ. Grenoble Alpes, CNRS, IPAG, 38000 Grenoble, France 
  \label{IPAG}
  \and
  Centro de Astrobiolog\'ia (CSIC-INTA), Torrej\'on de Ardoz, 28850 Madrid, Spain
  \label{CAB}
  \and  
  Instituto de Radioastronom\'ia Milim\'etrica (IRAM), Granada, Spain
  \label{IRAME}
  \and 
  LERMA, Observatoire de Paris, PSL Research University, CNRS, Sorbonne Universit\'e, UPMC, 75014 Paris, France  
  \label{LERMA}
  \and
  School of Earth and Space Exploration and Department of Physics, Arizona State University, Tempe, AZ 85287, USA
  \label{Arizona}
  \and
  Univ. Toulouse, CNRS, IRAP, 9 Av. du colonel Roche, BP 44346, 31028, Toulouse, France
  \label{IRAP}
  \and 
  Laboratoire de Physique de l’\'Ecole Normale Sup\'erieure, ENS, PSL Research University, CNRS, Sorbonne Universit\'e, Universit\'e de Paris, 75005 Paris, France 
  \label{ENS}
  \and
  Department of Physics and Astronomy, University of Pennsylvania, 209 South 33rd Street, Philadelphia, PA, 19104, USA
  \label{Pennsylvanie}
  \and 
  Institut d’Astrophysique de Paris, Sorbonne Universit\'e, CNRS (UMR7095), 75014 Paris, France
  \label{IAP}
  \and 
  Kavli Institute for Astrophysics and Space Research, Massachusetts Institute of Technology, Cambridge, MA 02139, USA
  \label{MIT}
  \and
  Caltech, Pasadena, CA 91125, USA
  \label{Caltech}
}

\abstract{%
In the past decade filaments have been recognised as a major structural element of the interstellar medium, the densest of these filaments hosting the formation of most stars. In some star-forming molecular clouds converging networks of filaments, also known as hub filament systems, can be found. These hubs are believed to be preferentially associated to massive star formation. As of today, there are no metrics that allow the systematic quantification of a filament network convergence. Here, we used the IRAM 30m NIKA2 observations of the Galactic plane from the GASTON large programme to systematically identify filaments and produce a filament convergence parameter map. We use such a map to show that: i. hub filaments represent a small fraction of the global filament population; ii. hubs host, in proportion, more massive and more luminous compact sources that non-hubs; iii. hub-hosting clumps are more evolved that non-hubs; iv. no discontinuities are observed in the properties of compact sources as a function of convergence parameter. We propose that the rapid global collapse of clumps is responsible for (re)organising filament networks into hubs and, in parallel,  enhancing the mass growth of compact sources. 
}
\maketitle
\section{Introduction}
\label{intro}
The existence of interstellar filaments has been known for more than four decades \cite{SE1979}. Their ubiquity within the interstellar medium, however, has only been recognised with the launch in 2009 of the {\it Herschel} mission, and its mapping of the sky in the far-infrared \cite{andre2010,molinari2010}. Since then filaments are front and centre of a large fraction of star formation research, and numerous breakthroughs have been made. It has been shown, for instance, that the majority of prestellar cores, and, by extension, the majority of stars form as the result of the fragmentation of individual 0.1pc-wide self-gravitating filaments \cite[e.g. ][]{arzoumanian2011,polychroni2013}.  Massive star-forming cores, on the other hand, seem to preferentially find themselves at the centre of  converging networks of filaments, or hubs \cite[e.g.][]{myers2009,peretto2013}.  This raises a number of questions on the physical relationship between filament convergence and the evolution of star-forming clouds. It has been proposed that hub formation is a two step process, with the formation of individual filaments first and then the collision of filament pairs in a second step \cite{kumar2020}. One key aspect of such scenario is the existence of two distinct formation path for individual and hub filaments. In the study presented here, we use the very sensitive 1.15mm continuum data of the GASTON large programme on the IRAM 30m to determine the time evolution of compact millimetre sources as a function of the local filament network convergence. 
 
\section{GASTON observations and catalogue}

GASTON is a 200h guaranteed-time large programme on the IRAM 30m telescope (PI: N. Peretto). About a third of the allocated time has been used to map a $\sim2$ deg$^2$ region of the Galactic plane around $\ell=24^{\rm{o}}$ at 1.15mm and 2mm using the NIKA2 instrument \cite{adam2018,perotto2020}. In the present study, we only make use of the 1.15mm image which has an angular resolution of $12''$, a pixel size of $3''$, and a rms sensitivity of 3.45~mJy/beam (see Fig.~\ref{flux_conv}), achieved with only half of the allocated time. In addition, we also make use of the GASTON catalogue of 1.15mm compact source presented in \cite{rigby2021}. In that catalogue, the distances and infrared brightness of all $\sim1400$ sources are given. The latter provides an indication on how bright the source parent clump is at 8$\mu$m, and is used here as a time evolution tracer \cite{rigby2021,watkinsprep}. Finally, all GASTON sources have been  cross-correlated with the Hi-GAL source catalogue \cite{molinari2016}.

\section{Quantifying filament convergence and hub definition}

\begin{figure}[t]
\begin{center}
\includegraphics[scale=0.45]{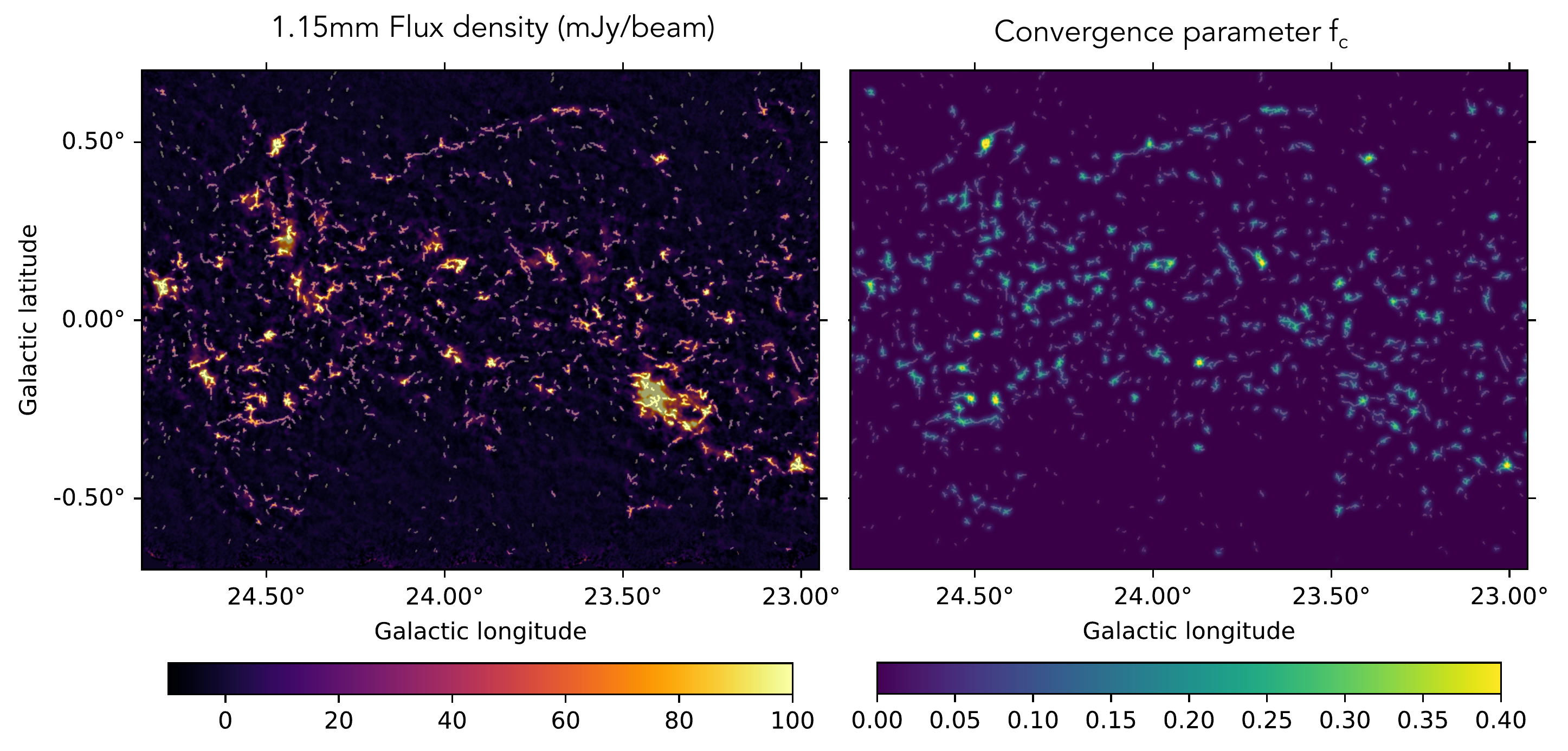}
\vspace{-0.7cm}
\caption{(left): GASTON 1.15mm image of the $\ell=24^{\rm{o}}$ region (colour scale). The identified filament skeleton is overlaid on top as grey solid lines. (right): Map of the convergence parameter $f_c$. Same as in the left hand side panel, the skeleton is overlaid on top of the image.
\label{flux_conv}}
\end{center}
\end{figure}

\begin{figure}[t]
\begin{center}
\includegraphics[scale=0.38]{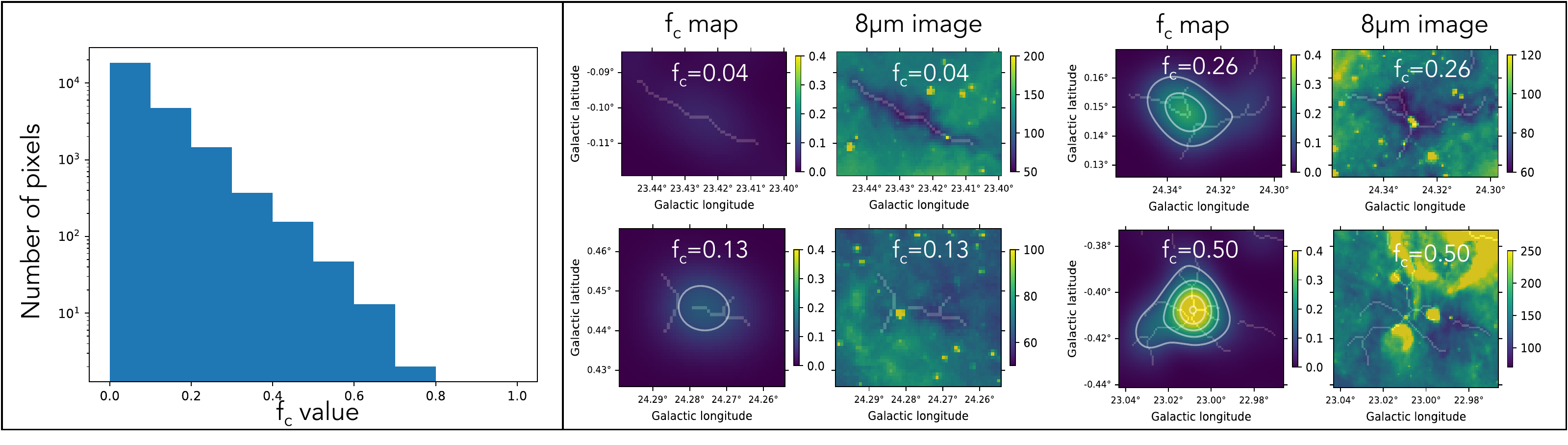}
\vspace{-0.5cm}
\caption{(left): Histogram of the $f_c$ values presented in Fig.~\ref{flux_conv}. (right): zoom-ins of 4 different sources exhibiting the range of $f_c$ values with their peak value indicated on the image. For each of the four pairs of images, we display the convergence map on the left and the corresponding {\it Spitzer} 8$\mu$m image on the right hand side. The skeletons are overlaid on top of each panel, and $f_c$ contours from 0.1 to 0.5 by step of 0.1 are also displayed on the convergence images. For context, the search radius $r_s$ is about $0.01^{\rm{o}}$.
\label{histo_zoom}  }
\end{center}
\end{figure}

Even though the study of hub filament systems has become, in the past 10 years, relatively widespread, there is no clear definition of what a hub is. For the first time, we propose to quantify the convergence of filaments in star-forming regions through the construction of a new convergence parameter, $f_c$. In order to do this, we first need to build the skeleton of the filament network within the GASTON 1.15mm image of the $\ell=24^{\rm{o}}$ region. For that purpose, we use the same second derivative method as presented in \cite[e.g. ][]{orkisz2019}. Skeletons are obtained by first using a local thresholding on the eigenvalue images, selecting only regions that have negative eigenvalues lower than -3 times the local standard deviation. The python {\it skeletonize} function is then used to thin each region to a 1-pixel width skeleton. We set a lower limit on the length of individual filaments to twice the beam size ($\sim 24''$ - 0.6pc at 5.2~kpc, the median cloud distance in the  $\ell=24^{\rm{o}}$ field). Figure \ref{flux_conv} shows the resulting skeleton overlaid on top of the GASTON 1.15mm image. A total of $\sim 2600$ filaments are identified in the $\ell=24^{\rm{o}}$ field. Compared to published catalogues of Galactic plane filaments, the GASTON filament density is larger, however those studies were typically geared towards the detection of larger-scale filaments \cite[e.g. ][]{schisano2020,li2016}. 
Since hubs have been in the past mostly visually identified in mid-infrared extinction, a visual check between our GASTON skeleton and the {\it Spitzer} 8$\mu$m image of the field is performed (see Fig.~\ref{histo_zoom}). The agreement is excellent and shows the power of our GASTON 1.15mm image in identifying small scale filamentary structures. 

Next we want to construct a new metric allowing to quantify how hub-like a network of filaments really is. This is the first ever attempt to provide an objective definition to what a hub is. For that purpose we define the following convergence parameter $f_c$:

\begin{equation}
f_c(x,y)=N_{\rm{fil}}\frac{\sum_{i=1}^{N_{\rm{pix}}}\cos(\Delta\theta)}{C_n}
\end{equation}

\noindent where $(x,y)$ is the pixel coordinate in the image, $N_{\rm{fil}}$ is the number of unique filaments that enter the search radius $r_{s}$, $N_{\rm{pix}}$ is the total number of skeleton pixels that are within the search radius, $\Delta\theta$ is the angle difference between the direction defined by the pixel $(x,y)$ and the skeleton pixel $i$, and the tangential direction of the skeleton at pixel $i$. Finally, $C_n$ is a normalisation constant such that a hub of six perfectly radial parsec-long connected filaments has a $f_c$ value of 1. The presence of the $N_{\rm{fil}}$ is needed to ensure that, for a fixed value of $N_{\rm{pix}}$, networks of converging filaments have larger $f_c$ values than single filaments. This definition of the convergence parameter implies that high $f_c$ value positions exhibit a high level of filament convergence within a radius $r_s$, while  low$f_c$ value implies the opposite. In this study we chose $r_s=39''$, which corresponds to $\sim1$~pc at the cloud median distance of 5.2~kpc. Figure \ref{flux_conv}(right) shows the resulting convergence map obtained from the underlying filament skeleton where high convergence spots can clearly be seen.

Figure \ref{histo_zoom}(left) shows the histogram of $f_c$ values presented in Fig.~\ref{flux_conv}, ranging from 0 to 0.75. One can see that the distribution can be described by a single power law, with no obvious breaks. Therefore the segmentation of clumps between hubs and non-hubs is somehow arbitrary. However one can still investigate what $f_c$ values typical hub filament systems have in order to separate hubs from non-hubs. Figure \ref{histo_zoom} shows four sources with increasing peak $f_c$ values and increasing filament complexity, the last two of which have peak $f_c$ values of 0.26 and 0.50 and exhibit a typical hub morphology.  We therefore decide to define hubs as connected groups of pixels with $f_c\ge0.2$.  By doing so, we identify 63 hubs and we find that only 5\% of skeleton pixels are located within hubs. Hub filament systems are rare objects.

\section{Source properties as a function of convergence parameter}
 \begin{figure}[t]
\begin{center}
\includegraphics[scale=0.41]{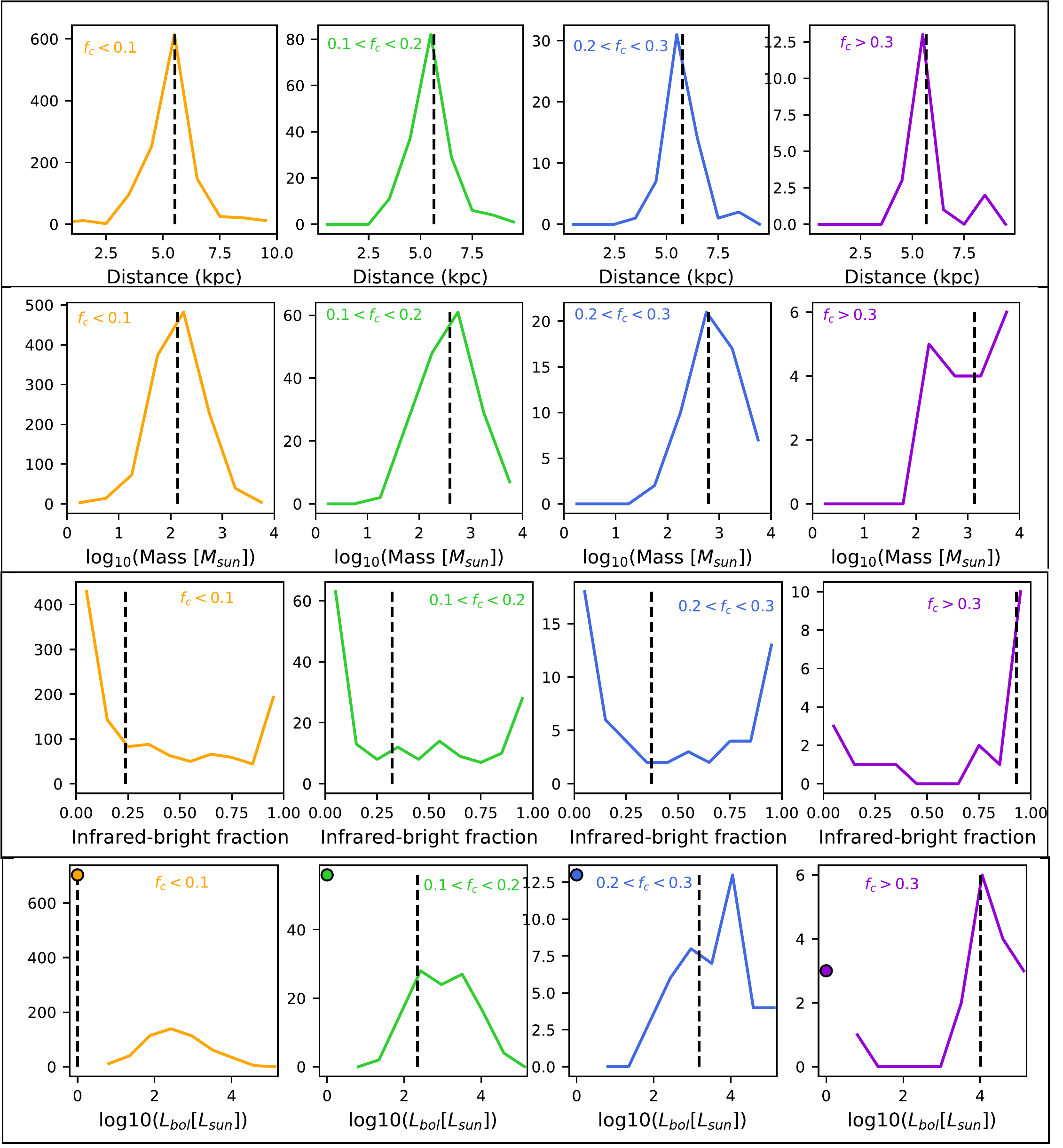}
\caption{(1$^{\rm{st}}$ row): Histogram of distances for the GASTON 1.15mm compact sources identified by \cite{rigby2021}. Each panel corresponds to a range of $f_c$ values indicated in each panel. The vertical dashed lines show the median values. (2$^{\rm{nd}}$ row): same as top row but for the masses of the compact sources. (3$^{\rm{rd}}$ row): same as top row but for the infrared darkness parameter. (4$^{\rm{th}}$ row): Same as top row but for the 70$\mu$m-based bolometric luminosities. The circular symbols indicate the number of sources with no Hi-GAL 70$\mu$m sources, i.e. sources for which the bolometric luminosity could not be computed.
\label{prop} }
\end{center}
\end{figure}

Now that we have established a working definition of a hub, we can investigate the properties of compact GASTON sources as a function of the type of clump they find themselves into, and more generally, as a function of their convergence parameter. For that purpose we use the catalogue of compact sources identified by \cite{rigby2021} and split the $\sim1400$ compact 1.15mm GASTON sources in four $f_c$ bins: $f_c<0.1$,   $0.1<f_c<0.2$, $0.2<f_c<0.3$, and $f_c>0.3$, in which there are 1186, 170, 56 and 19 compact sources, respectively. Based on this splitting, we can already see that only 8\% of GASTON compact sources are embedded within a hub. 

First, we want to make sure that there is no distance bias when splitting out source sample into $f_c$ bins. Figure~\ref{prop}(1$^{\rm{st}}$ row) shows the distance histograms for each $f_c$ bin. The vertical dash line shows the median value. One can see that there is nearly no change whether in the shape of the distribution or in the median distance values, all at about 5.2~kpc. Then we look at the distribution of the  source masses. Figure~\ref{prop} (2$^{\rm{nd}}$ row) clearly shows that there is an increase in source mass as the convergence parameter increases. In fact the median mass for the $f_c<0.1$ bin is at 135~M$_{\odot}$ while that of the $f_c>0.3$ bin is at 1350~M$_{\odot}$, an order of magnitude larger. This shows that a larger fraction of compact sources associated with hubs are massive, although massive compact sources are also found in low $f_c$ regions.

Next, we investigate how the infrared-bright fraction ($f_{\rm{IRB}}$) of the GASTON sources evolves as a function of the convergence parameter. The brightness parameter is a measurement of how bright a cloud is as 8$\mu$m. For instance, infrared dark clouds that do not host any star formation activity yet will have an infrared brightness parameter of 0. The first of the four clumps shown in Fig.~\ref{histo_zoom} is a good example of such pristine structure. On the other hand a cloud which is bright at 8$\mu$m, such as the last one of the four clumps presented in Fig.~\ref{histo_zoom}, will have an infrared brightness close to 1. Infrared brightness is believed to be a good tracer of clump evolution \cite{rigby2021,watkinsprep}. In Fig.\ref{prop}(3$^{\rm{rd}}$  row) we show the histograms of infrared brightness for the same four convergence parameter bins. There, we see a continuous shift from clumps that are mostly infrared dark at low $f_c$ values, to clumps that are mostly infrared bright at large $f_c$ values. One can interpret this in two ways: hubs are the late stages of clump evolution in which filament convergence increases with time; infrared dark hubs have a very short lifetime and become very quickly infrared bright via a strong star formation activity. Note that these two scenarios are not mutually incompatible. A similar picture can be drawn from Fig. \ref{prop}(4$^{\rm{th}}$  row) where the distribution of bolometric luminosities are presented. Here, the luminosities are being computed from the 70$\mu$m flux of internal protostellar sources. On that figure we can clearly see that hubs host the formation of  more luminous sources, including all sources with $L_{\rm{bol}}\le10^5$L$_{\odot}$, a result that was already found by \cite{kumar2020}.


 \section{Conclusion}
 
In this study, we have  provided, for the first time, a quantitative definition of what a hub is, and shown that they in fact represent a small fraction of the filament population. We have also shown that hubs have a higher proportion of massive, luminous, and more evolved compact sources within them. Also, in all distributions of properties we have investigated, we do not observe any discontinuity as a function of the convergence parameter but rather continuous and smooth transitions from single filaments to highly convergent filament networks. This is a clear indication that hubs do not represent a different population of filament with their specific formation scenario. Instead we propose that it is the rapid global collapse of clumps that is responsible for (re)organising  filament networks into hubs and, in parallel, for leading to the mass growth of compact sources.

\section*{Acknowledgements} \label{ack}
NP and AJR would like to thank the STFC for financial support under the consolidated grant numbers ST/N000706/1 and ST/S00033X/1, and the Royal Society for providing computing resources under Research Grant number RG150741. We would like to thank the IRAM staff for their support during the campaigns. The NIKA2 dilution cryostat has been designed and built at the Institut N\'eel. In particular, we acknowledge the crucial contribution of the Cryogenics Group, and in particular Gregory Garde, Henri Rodenas, Jean Paul Leggeri, Philippe Camus. This work has been partially funded by the Foundation Nanoscience Grenoble and the LabEx FOCUS ANR-11-LABX-0013. This work is supported by the French National Research Agency under the contracts "MKIDS", "NIKA" and ANR-15-CE31-0017 and in the framework of the "Investissements d’avenir” program (ANR-15-IDEX-02). This work has benefited from the support of the European Research Council Advanced Grant ORISTARS under the European Union's Seventh Framework Programme (Grant Agreement no. 291294). F.R. acknowledges financial supports provided by NASA through SAO Award Number SV2-82023 issued by the Chandra X-Ray Observatory Center, which is operated by the Smithsonian Astrophysical Observatory for and on behalf of NASA under contract NAS8-03060. 
%
%
%

\end{document}